\documentclass[12pt]{iopart}
\usepackage{graphicx}
\usepackage{bm}
\usepackage[numbers,sort&compress]{natbib}
\usepackage{epstopdf}

\usepackage[final,bookmarks=true,bookmarksopen=true,bookmarksdepth=1,bookmarksnumbered=true,pdfstartview=FitH,pdfborder={0 0 0}]{hyperref}

\def\equationautorefname~#1\null{%
	Eq.~(#1)\null
}

\renewcommand{\vec}[1]{
	\mathbf{#1}
}

\begin{document}

\date{\today}

\title{Dirac model of electronic transport in graphene antidot barriers}

\author{M R Thomsen, S J Brun and T G Pedersen}

\address{Department of Physics and Nanotechnology, Aalborg University, Skjernvej 4A, DK-9220 Aalborg \O st, Denmark and}
\address{Center for Nanostructured Graphene (CNG), DK-9220 Aalborg \O st, Denmark}
\ead{mrt@nano.aau.dk}

\begin{abstract}
In order to use graphene for semiconductor applications, such as transistors with high on/off ratios, a band gap must be introduced into this otherwise semimetallic material. A promising method of achieving a band gap is by introducing nanoscale perforations (antidots) in a periodic pattern, known as a graphene antidot lattice (GAL). A graphene antidot barrier (GAB) can be made by introducing a 1D GAL strip in an otherwise pristine sheet of graphene. In this paper, we will use the Dirac equation (DE) with a spatially varying mass term to calculate the electronic transport through such structures. Our approach is much more general than previous attempts to use the Dirac equation to calculate scattering of Dirac electrons on antidots. The advantage of using the DE is that the computational time is scale invariant and our method may therefore be used to calculate properties of arbitrarily large structures. We show that the results of our Dirac model are in quantitative agreement with tight-binding for hexagonal antidots with armchair edges. Furthermore, for a wide range of structures, we verify that a relatively narrow GAB, with only a few antidots in the unit cell, is sufficient to give rise to a transport gap.
\end{abstract}


\section{Introduction}

Graphene has been the subject of intense research since its discovery in 2004 \cite{novoselov2004electric}. Especially the ultrahigh mobility \cite{du2008approaching,bolotin2008ultrahigh,bolotin2008temperature} of pristine graphene makes it a promising platform for novel nanoelectronic devices. Pristine graphene does not have a band gap, which makes it ill-suited for semiconductor applications, such as transistors with high on/off ratios for logic applications. 
Band gaps in graphene have been demonstrated experimentally in graphene nanoribbons \cite{han2007energy}, gated bilayer graphene \cite{zhang2009direct,xia2010graphene} and patterned adsorption of hydrogen on graphene \cite{balog2010bandgap}. Another promising method for creating a tunable band gap in graphene is by introducing nanoscale perforations in a periodic fashion, known as graphene antidot lattices (GALs) or graphene nanomeshes \cite{pedersen2008graphene,furst2009density,furst2009electronic}. The magnitude of the band gap depends on the size of the antidots, size of the unit cell and on edge chirality \cite{brun2014electronic,pedersen2008graphene,petersen2009quasiparticle,vanevic2009character,trolle2013large}. It has been shown that the band gap of GALs with relatively small antidots follows a simple scaling rule proposed by Pedersen \textit{et al.} \cite{pedersen2008graphene}.

Several methods have been used to produce GALs experimentally, including \mbox{e-beam} lithography \cite{eroms2009weak,giesbers2012charge,xu2013controllable}, diblock copolymer templates \cite{bai2010graphene,kim2010fabrication,kim2012electronic}, anodic aluminum oxide templates \cite{zeng2012fabrication}, nanosphere lithography \cite{wang2013cvd} and nanoimprint lithography \cite{liang2010formation}. The antidots range in size between a few nanometers and several hundred nanometers, depending on the fabrication method. The antidots synthesized with these methods are often round, but it has been demonstrated experimentally that armchair and zigzag edges in GALs are stable and can be synthesized selectively \cite{jia2009controlled,girit2009graphene,oberhuber2013weak}. Recent experimental studies of transport in GALs have shown on/off ratios in the range between 4 and 100~\cite{zeng2012fabrication,liang2010formation,bai2010graphene}. These values are still too low for logic applications~\cite{schwierz2010graphene}, but the results are important indicators that devices based on GALs could be used to make efficient transistors. 
The electronic transport properties of GALs have also been studied theoretically. The transport through graphene antidot barriers (GABs), i.e.\ 1D periodic antidot structures in an otherwise pristine sheet of graphene, has previously been studied for small systems using a tight-binding (TB) formalism~\cite{pedersen2012transport,gunst2011thermoelectric}. These studies showed that just a few antidots in the unit cell of the GAB is sufficient to suppress the transport within the band gap region. Furthermore, Berreda \textit{et al.}~\cite{berrada2013graphene} have simulated three different graphene field-effect transistors based on GALs with band gaps of about 500 meV. They showed that their simulated devices had on/off ratios as high as 7400, which is close to that of silicon based MOSFETs that have on/off ratios on the order of $10^4$ to $10^7$~\cite{schwierz2010graphene}.

Experimentally feasible GALs are typically too large to handle with traditional atomistic models, such as TB and DFT. However, models based on the Dirac equation (DE) are in the continuum regime and are therefore able to handle arbitrarily large structures.
In this paper, we will use the Dirac equation (DE) with a spatially varying mass term to calculate the scattering of Dirac electrons in GABs that are periodic in one dimension. It has previously been shown that the DE on this form can be used to calculate the band structure of GALs \cite{brun2014electronic,furst2009electronic}. In addition, the DE has previously been used to calculate scattering of Dirac electrons on a single circular mass barrier \cite{masir2011scattering}, a single circular electrostatic barrier \cite{heinisch2013mie} and simple barriers of constant and finite mass \cite{gomes2008tunneling}. The advantages of our approach are that it works for any antidot shape and for an arbitrary arrangement of antidots. Furthermore, our method can easily be extended to a 1D periodic case. Another advantage that arises from using the DE is that all results are scalable, i.e., the results are invariant when all lengths are scaled up by some factor and all energies are scaled down by the same factor. We use a Green's tensor area integral equation method (AIEM) in order to solve the DE. We will focus on the transport of a plane electron wave through GABs with two different types of hexagonal antidots, namely antidots with zigzag edges and antidots with armchair edges. We compare the results of our Dirac model with results obtained with TB.

\section{Theory and methods}

\begin{figure}[htb]
\centering
\includegraphics[width=0.5\columnwidth]{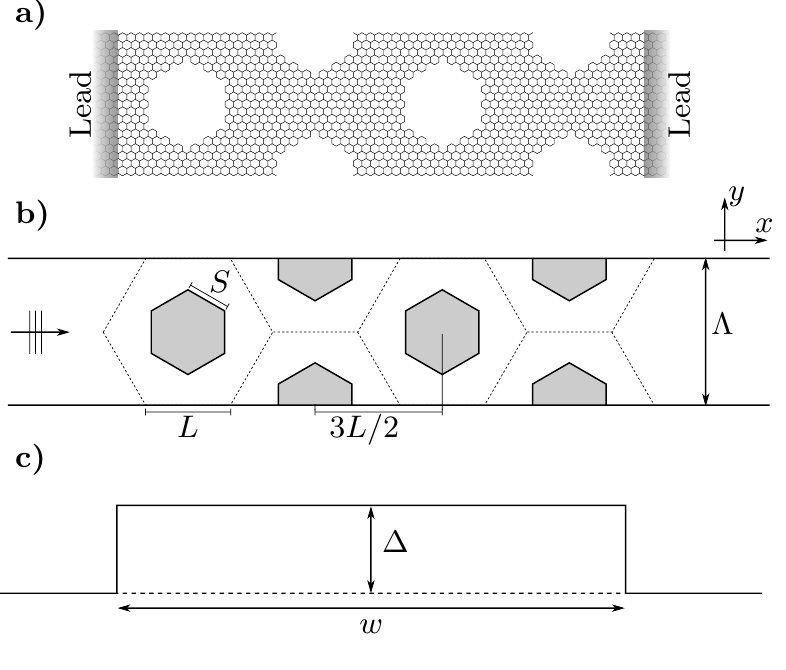}
\caption{Unit cells of a GAB with four rows of armchair antidots. a) Geometry used in TB. b) Geometry used with the DE, where the mass term has a constant value of $\Delta_0$ inside the shaded areas and is vanishing elsewhere. c) Dirac mass barrier (DMB) with height $\Delta$ and width $w$.}
\label{fig:structure}
\end{figure}

In this section, we will set up a Green's tensor AIEM to calculate the scattering of Dirac electrons on arbitrary graphene antidot structures, where an electron wave is incident on the structure and the resulting total wave function is calculated. Once this general method has been set up, we will specialize to scattering of Dirac electrons on GABs as shown in \autoref{fig:structure}b. The idea of using an AIEM to solve inhomogeneous differential equations is not new. In fact, it has been used extensively to solve scattering problems in optics \cite{sondergaard2007modeling,novotny2006principles} and we utilize several of the same techniques to calculate scattering of electron waves in graphene.

The DE for a graphene sheet with a spatially varying mass term $\Delta(\vec{r})$ has the form~\cite{furst2009electronic}

\begin{equation}
(v_F\bm{\sigma}\cdot \vec{p}+\Delta(\vec{r})\sigma_z-\bm{I}E)\Psi = 0\, ,\label{eq:Dirac1}
\end{equation}
\begin{equation}
(v_F\bm{\sigma}\cdot \vec{p}-\Delta(\vec{r})\sigma_z-\bm{I}E)\Psi' = 0\, ,
\end{equation}

\noindent where $\Psi = \{\psi_A, \psi_B\}$ and $\Psi' = \{\psi'_B, \psi'_A\}$ are the wave functions associated with the $K$ and $K'$ valleys, respectively, $\bm{\sigma} = \{\sigma_x,\sigma_y\}$ and $\sigma_z$ are the Pauli matrices, $\bm{p} = \{\hat{p}_x,\hat{p}_y\}$ is the momentum operator, and $v_F$ is the Fermi velocity. The mass term has a constant value of $\Delta_0$ inside the antidots and is vanishing elsewhere. This effectively makes electrons massive inside the antidots, making it energetically unfavorable to enter them. The mass term should be sufficiently large in order to model actual holes in graphene. It should generally be much larger than the electron energy $\Delta_0 \gg |E|$. We use $\Delta_0 = 170\mbox{ eV}/L$ in all our calculations, which is identical to the value used in Ref.~\cite{brun2014electronic}. Due to the similarity of the $K$ and $K'$ equations, we can restrict our analysis to one of them. 
The wave function of the incident wave $\Psi_0$ must be a solution to the case without a scatterer (pristine graphene), i.e.\ the case where $\Delta(\vec{r})=0$ everywhere. This simply reduces \autoref{eq:Dirac1} to the DE without a mass term.
We will use incident plane waves on the form $\Psi_0 = 2^{-1/2}(1,e^{i\varphi})^Te^{i\vec{k}\cdot\vec{r}}$, where $\varphi$ is the polar angle of $\vec{k}$.

The Green's tensor $\bm{G}$ between an observation point $\vec{r}$ and a source point $\vec{r}'$ is defined as the solution to the equation

\begin{equation}
(v_F\bm{\sigma}\cdot \vec{p}-\bm{I}E)\bm{G}(\vec{r},\vec{r}') = -\bm{I}\delta(\vec{r}-\vec{r}')\, .
\end{equation}

\noindent The solution must obey the radiation condition, which states that the solution should asymptotically tend towards an outward propagating wave proportional to $e^{ikr}/\sqrt{kr}$. This uniquely specifies the Green's tensor as

\begin{equation}
\bm{G}(\vec{r},\vec{r}') = \frac{k}{4i}\left(\begin{array}{cc} H_0^{(1)}(kr) & -ie^{-i\theta}H_1^{(1)}(kr)\\
-ie^{i\theta}H_1^{(1)}(kr) & H_0^{(1)}(kr)\end{array}\right)\,,
\end{equation}

\noindent where $H_n^{(1)}$ is the $n$'th order Hankel function of the first kind, $k = E/\hbar v_F$, $r = |\vec{r}-\vec{r'}|$ and $\theta$ is the polar angle of  $\vec{r}-\vec{r}'$. By subtracting the DE without a mass term from \autoref{eq:Dirac1} we get

\begin{equation}
(v_F\bm{\sigma}\cdot \vec{p}-\bm{I}E)(\Psi-\Psi_0) = -\Delta(\vec{r})\sigma_z\Psi\,,
\label{eq:Dirac4}
\end{equation}

\noindent which has the solution

\begin{equation}
\Psi(\vec{r}) = \Psi_0(\vec{r}) + \int \widetilde{\Delta}(\vec{r}')\sigma_z\bm{G}(\vec{r},\vec{r}')\Psi(\vec{r}') d^2r'\,,
\label{eq:Dirac2}
\end{equation}

\noindent where $\widetilde{\Delta}(\vec{r}) = \Delta(\vec{r})/\hbar v_F$. This is the central equation for the Green's tensor AIEM, which we will use to calculate the scattering of Dirac electrons on antidot structures. It can be demonstrated that the equation is invariant when all lengths are scaled up by some factor and all energies and mass terms are scaled down by the same factor. This effectively means that the computational time is scale invariant.

The main advantage of this approach is that we only need to consider points $\vec{r'}$ where $\Delta(\vec{r}') \neq 0$, i.e.\ inside the antidot. Once we know the wave function inside the antidot, it is a simple matter to use \autoref{eq:Dirac2} to calculate the wave function at any other position. We solve this self-consistently by discretizing the area inside the antidots into a number of small areas $\delta A_{i}$ with centers $\vec{r}_{i}$. The integral is then completed by assuming that the mass term and the wave function are constant inside each area element and by approximating the Green's tensor between element $i$ and $j$ as

\begin{equation}
\bm{G}_{ij} \simeq \left\{ \begin{array}{cc}
(\delta A_j)^{-1} \int_{\delta A_j} \bm{G}(\vec{r}_i,\vec{r}') d^2 r' & \quad\mbox{if } i=j\\
\bm{G}(\vec{r}_i,\vec{r}_j) & \quad\mbox{if } i\neq j
\end{array} \right. \,.
\end{equation}

\noindent The self-interaction element $i=j$ may be calculated by approximating the area element with a circle, with radius $r_{eq}$, of equivalent area, i.e.\ $\delta A_j = \pi r_{eq}^2$, and integrating in polar coordinates

\begin{equation}
\bm{G}_{ii} \simeq [1/(\pi r_{eq}^2 k) - i H_1^{(1)}(kr_{eq})/(2r_{eq})]\bm{I} \,.
\end{equation}

\noindent We now have all the ingredients necessary to solve the scattering problem. It is then a simple matter of using matrix inversion or some efficient  iterative scheme to solve for the wave function inside the antidots.

We will specialize to the case of GABs, where the antidot structure is periodic along the $y$-direction with period $\Lambda$ as shown in \autoref{fig:structure}b. We will focus on hexagonal antidots arranged in a GAL configuration, meaning that the antidot lattice vectors are parallel to the carbon-carbon bonds of the graphene lattice as shown in \autoref{fig:structure}a. The antidots are chosen, such that they have either armchair edges (denoted armchair antidots) or zigzag edges (denoted zigzag antidots). The structures are described by the side length $L$ of the GAL unit cell, the side length $S$ of the antidot and the number of antidots $N$ in the GAB unit cell, see \autoref{fig:structure}b. All distances are in units of the graphene lattice constant $a$. The notation $N-$A\{$L,S$\}GAL and $N-$Z\{$L,S$\}GAL will be used to describe barriers with $N$ armchair and zigzag antidots, respectively, in GAL a configuration.

In the periodic case, the scattered part of the wave function is given by an infinite sum of integrals over unit cells. By shifting all integrals to the zeroth unit cell, we can take the sum inside the integral and, thus, only integrate over the area of the zeroth unit cell $A_0$. All shifted wave functions are related to the wave function in the zeroth unit cell by the Bloch condition $\Psi(\vec{r}+m\Lambda \hat{y}) = \Psi(\vec{r})e^{imk_y\Lambda}$, where $m$ is an integer, $\Lambda=\sqrt{3}L$ is the period, $k_y=k\sin(\varphi)$ and $\varphi$ is the angle of incidence. We may then write the wave function as

\begin{equation}
\Psi(\vec{r}) = \Psi_0(\vec{r}) + \int_{A_0} \widetilde{\Delta}(\vec{r'})\sigma_z\widetilde{\bm{G}}(\vec{r},\vec{r}')\Psi(\vec{r}') d^2r'\,,
\end{equation}

 \noindent where $\widetilde{\bm{G}}$ is a modified Green's tensor given by

\begin{equation}
\widetilde{\bm{G}}(\vec{r},\vec{r}') = \sum_{m=-\infty}^\infty \bm{G}(\vec{r},\vec{r}'-m\Lambda\hat{y})e^{ik_ym\Lambda}\,.
\end{equation}

\noindent This sum is extremely slowly convergent. However, once it has been determined, the problem of finding the wave function is no harder than in the non-periodic case. Using Graf's theorem \cite{abramowitz1965handbook} the Green's tensor may be restated as

\begin{equation*}
\widetilde{\bm{G}}(\vec{r},\vec{r}') = \sum_{m=-M}^M\bm{G}(\vec{r},\vec{r}'-m\Lambda\hat{y})
\end{equation*}
\begin{equation}
\quad\quad+ \frac{k}{4i}\sum_{n=-\infty}^\infty i^nJ_n(kr)e^{-in\theta} \left(\begin{array}{cc}
	S_n & -S_{n-1}\\
	-S_{n+1} & S_n
\end{array}\right)\,,
\end{equation}

\noindent where $J_n$ is the $n$'th order Bessel function of the first kind and $S_n$ is the $n$'th order \textit{lattice sum} given by

\begin{equation}
S_n = \sum_{m=M+1}^\infty H_n^{(1)}(km\Lambda)\left( e^{ik_ym\Lambda} + (-1)^n e^{-ik_ym\Lambda} \right)\,.
\end{equation}

\noindent We have taken the contribution of $M$ unit cells on either side of the zeroth unit cell outside the lattice sum as they may not satisfy the condition for using Graf's theorem. In fact, Graf's theorem is only satisfied when the largest distance between area elements within one unit cell is smaller than $(M+1)\Lambda$. Therefore, $M$ is chosen to be the smallest integer that satisfies this condition. The lattice sum is actually also extremely slowly convergent, but there are two advantages of writing $\widetilde{\bm{G}}$ using the lattice sum: 1) The lattice sum does not depend on the observation point, so it needs only be calculated once for a given choice of $k\Lambda$ and angle of incidence $\varphi$, and 2) it can be calculated efficiently using the integral representation described in Ref.~\cite{yasumoto1999efficient}.

The transmittance $T$ through the barrier is simply the transmitted current $I$ divided by the incident current $I_0$. The current is calculated by integrating the $x$-component of the current density over one period $I = \int_{uc} j_x\, dy$, where the current density is given by $j_x = \Psi^\dagger \hat{j}_{x} \Psi$ using the current density operator $\hat{j}_{x} = -ev_F\sigma_x$. In the limit of vanishing bias, the conductance is $G=G_0T$, where $G_0=2e^2/h$ is the conductance quantum.

The transport through a GAB may be approximated by replacing the actual structure with a simple barrier as shown in \autoref{fig:structure}c \cite{pedersen2012transport}. This type of barrier is referred to as a Dirac mass barrier (DMB). In this approach, we define the width of the barrier as $w=N(3L/2)$ and take the barrier height $\Delta$ to be half the band gap of the fully periodic case. We use the approximation of the band gap given by Eq.~A6 in Ref.~\cite{brun2014electronic}.

In order to assess the accuracy of our model, we compare our results with spectra calculated using the Landauer approach with a nearest-neighbor TB Hamiltonian, as outlined in Ref.~\cite{pedersen2012transport}. We use a hopping integral of $\gamma = 3.033$ eV and, for numerical stability, we add a small imaginary part to the energy, such that $E\rightarrow E+i\varepsilon$, where we use  $\varepsilon = 10^{-5}$ eV. In the Dirac models, we average over valleys in order to obtain the conductance per valley. All TB spectra are therefore divided by a factor of two in order to directly compare with the DE.

In order to make a quantitative comparison between the models for a wide range of structures, we define a transport gap using the lowest positive energy, at which the conductance rises above $G_0/2$. Due to exact electron-hole symmetry, the transport gap will then be twice this energy. Long zigzag edges give rise to very localized states in TB. However, in a real device with even a small amount of disorder, we do not expect these states to support electronic transport. This effect can be introduced heuristically by convolving the TB conductance spectra with a Gaussian or by using a larger imaginary part of the energy $i\varepsilon$. Therefore, in our calculations of the transport gap we convolve with a Gaussian having a full width at half maximum of $0.1 \mbox{ eV}/L$.

\section{Results}

In this section, we present the results of our Dirac model and compare them with TB and the DMB model. Results are presented for both armchair and zigzag antidots. The geometries used in our Dirac model are created such that the total area of the antidots equals the total area of the removed atoms of the corresponding structure used in TB.

\subsection{Armchair antidots}
\begin{figure}[htb]
\centering
\includegraphics{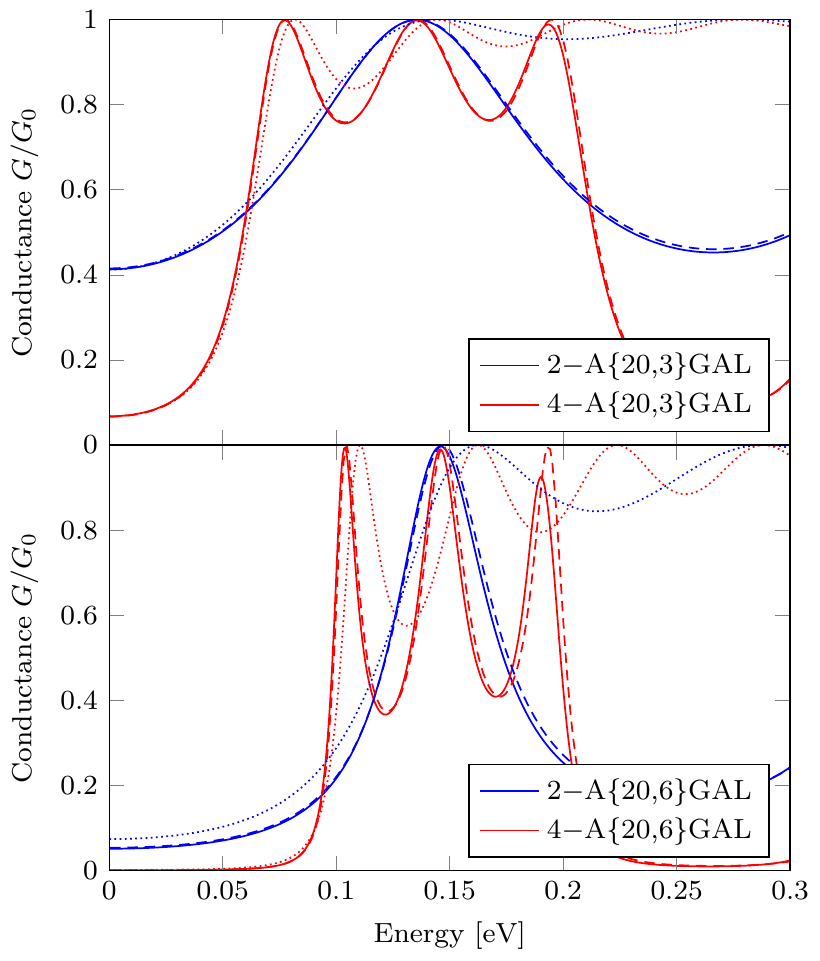}
\caption{Conductance of GABs with armchair antidots calculated with TB (solid), the DE (dashed), and the DMB model (dotted).}
\label{fig:transport_HA}
\end{figure}

We start out by considering GABs in GAL configuration containing armchair antidots. The conductance spectra of two different barriers with armchair antidots is shown in \autoref{fig:transport_HA}. The results are only displayed for positive energies, as the results for negative energies follow from exact electron-hole symmetry in the models. There is excellent agreement between TB and our Dirac model. The DMB model is in good agreement with the more sophisticated ones at low energies, but it deviates substantially at large energies. Furthermore, it is seen that there are always $N-1$ subpeaks in the conductance peak, which is consistent with previous calculations for graphene nanoribbons with antidot arrays \cite{zhang2010band}. This means that as the number of antidots in the unit cell increases, the subpeaks will come closer to each other and eventually merge into a single step-like plateau.

\begin{figure}[tb]
\centering
\includegraphics{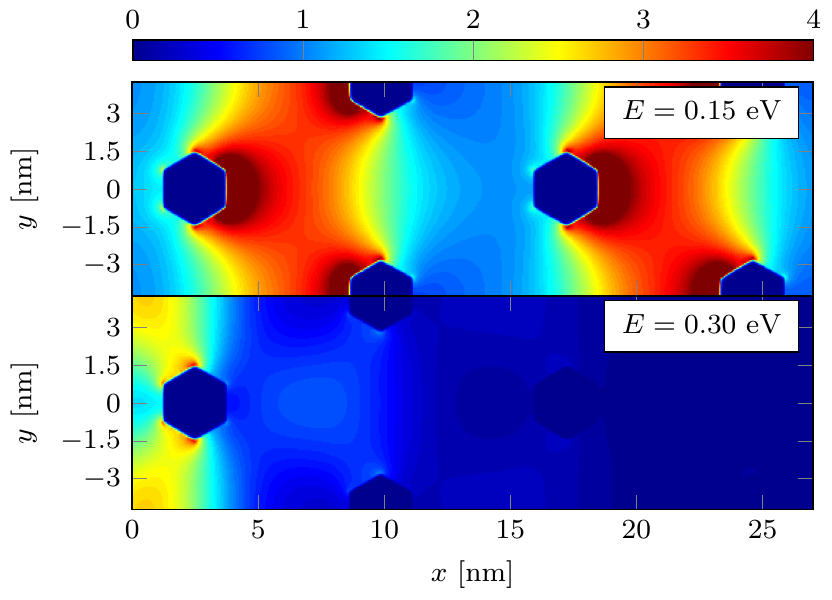}
\caption{Probability density of electrons in a $4-$A\{20,6\}GAL barrier calculated using the DE. The probability density is measured relative to the incident wave.}
\label{fig:probdens_HA}
\end{figure}

In order to gain insight into the electronic transport through a GAB, we compute the electron probability density for a $4-$A\{$20,6$\}GAL barrier at two different electron energies, as shown in \autoref{fig:probdens_HA}. The two lowest bands in the electronic band structure for the fully periodic structure have energies in the intervals [0.09; 0.24]~eV and [0.31; 0.50]~eV as given in Ref.~\cite{brun2014electronic}. We expect low conductance in the band gap regions of the fully periodic structure and high conductance elsewhere. At $E=0.15$~eV, the electron has an energy within the first band, and the probability density inside the barrier is therefore quite high, which results in a very high conductance of $G \simeq 0.91G_0$. However, at $E=0.3$ eV, the electron has an energy within a band gap, and the probability density inside the barrier is therefore rather low, which results in a much lower conductance of $G \simeq 0.02G_0$. This means that the conductance is low for energies at which the barrier region does not support any electron states.

\begin{figure}[htb]
\centering
\includegraphics{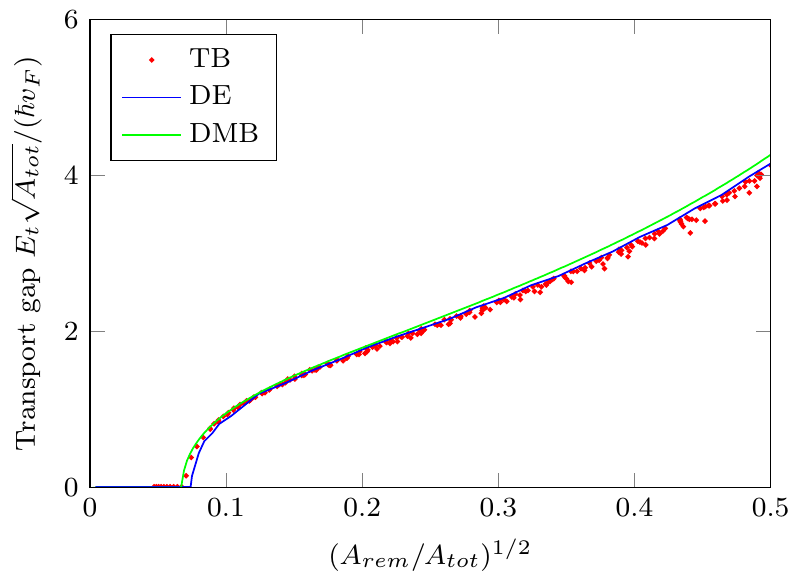}
\caption{Transport gap of $4-$A\{$L,S$\}GAL barriers calculated using TB, the DE and the DMB model.}
\label{fig:transportgap_HA}
\end{figure}

Armchair antidots do not support localized edge states, which means that, in the limit of very wide barriers, the transport gap should equal the band gap of the fully periodic structure. It is interesting, however, to see if a barrier with only a few antidots in the unit cell is able to block electron transport in the band gap region. Figure \ref{fig:transportgap_HA} shows the transport gap of a large range of GABs with just 4 antidots in the unit cell. In accordance with Ref.~\cite{brun2014electronic}, the results are scaled with the total area of the GAL unit cell $A_{tot} = 3\sqrt{3}L^2/2$ and the area of a single antidot $A_{rem}$. It is seen that the transport gap opens up for antidots with a size $(A_{rem}/A_{tot})^{1/2}>0.07$. The transport gap is exactly zero for small antidots, as the conductance is higher than $G_0/2$ at vanishing energy. The abrupt opening of the transport gap is due to the horizontal slope of the conductance as a function of energy at small energies. This means that as soon as the structure is large enough for the conductance at vanishing energy to fall below $G_0/2$, the transport gap will increase rapidly. The exact location of the onset of the transport gap is, thus, sensitive to the choice of transport gap definition. However, the remaining values are not too sensitive to the exact definition of the transport gap, since the slope of the conductance spectrum is typically very large near the transport gap. It is seen that both Dirac models are in excellent agreement with TB.

\subsection{Zigzag antidots}
\begin{figure}[tb]
\centering
\includegraphics{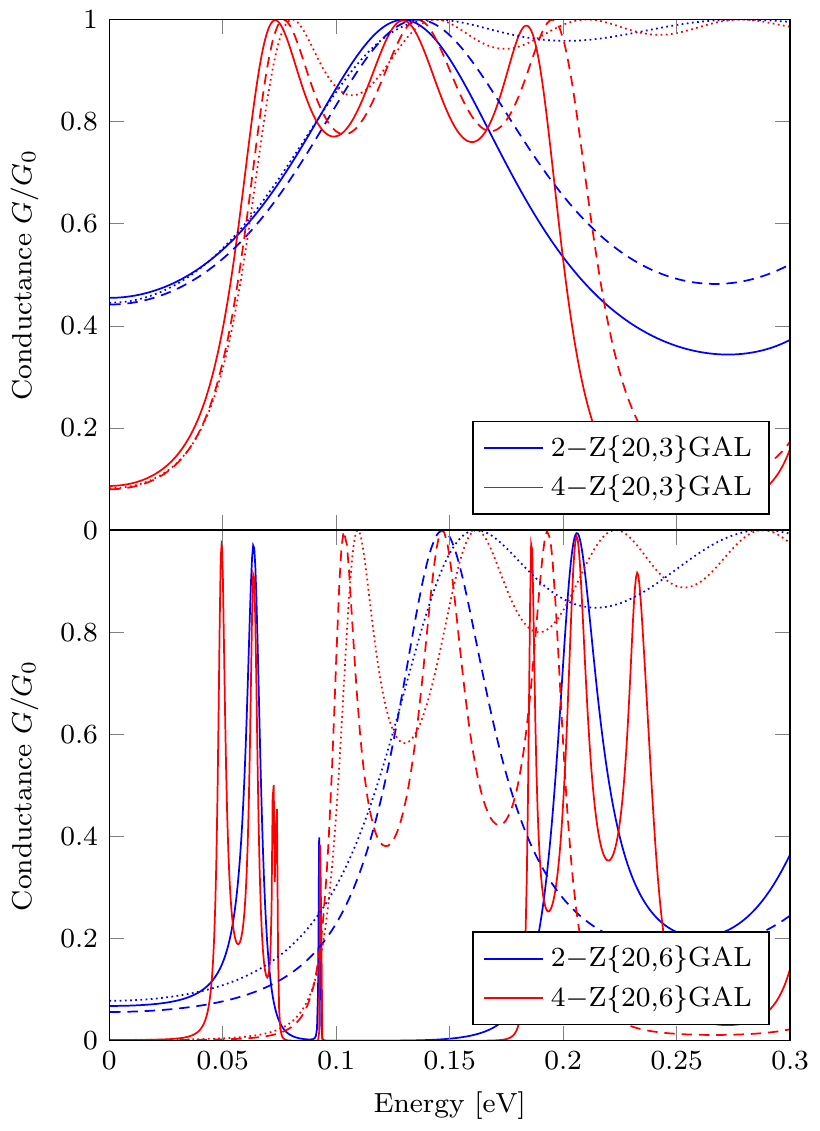}
\caption{Conductance of GABs with zigzag antidots calculated using TB (solid), the DE (dashed), and the DMB model (dotted).}
\label{fig:transport_HZ}
\end{figure}

Conductance spectra calculated with our Dirac model for two different barriers with zigzag antidots are shown in \autoref{fig:transport_HZ} and compared to TB. There is a fairly good agreement between the models for the smaller antidots, but the agreement is very poor for the larger structure. These deviations arise due to the highly localized state in the TB spectrum near $0.09$ eV, which is a result of the long zigzag edges. The deviations between TB and the DE have also been observed in the calculation of the band gap of fully periodic GALs \cite{brun2014electronic}.

\begin{figure}[tb]
\centering
\includegraphics{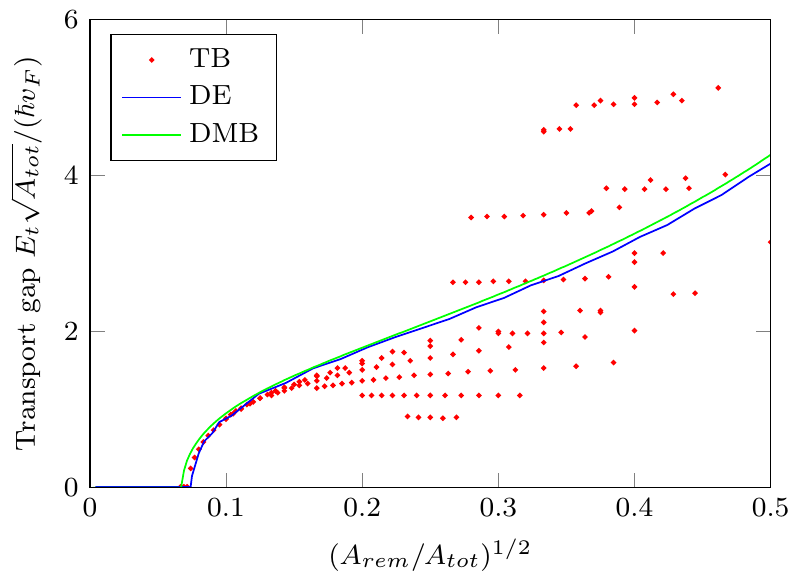}
\caption{Transport gap of $4-$Z\{$L,S$\}GAL barriers calculated using TB, the DE, and the DMB model.}
\label{fig:transportgap_HZ}
\end{figure}

It has previously been shown that the band gap shrinks substantially, compared to simple scaling laws, for structures with large zigzag antidots due to the presence of edge states \cite{brun2014electronic}. The shrinking of the band gap is only predicted by the TB model, as the DE with a mass term does not distinguish between zigzag and armchair edges. In the calculation of the transport gap, we overcome some of the effects of very localized edge states in our TB calculations by convolving all TB conductance spectra with a Gaussian. This smears out very narrow features of the conductance spectra, while preserving those that are not. We compare the transport gap calculated with the DE with those calculated with TB and the DMB model in \autoref{fig:transportgap_HZ}. The Dirac models are in fairly good agreement with TB for small antidots with size $(A_{rem}/A_{tot})^{1/2}<0.15$. However, for larger antidots, there is generally a poor agreement. This is again due to the presence of localized states in the TB spectra. An interesting aspect of the TB transport gap, however, is that it is generally much higher than the TB band gap, which is given in Ref.~\cite{brun2014electronic}. TB predicts that the band gap almost closes for large zigzag antidots, whereas the TB transport gap does not. In fact, the TB transport gap is often higher than the one predicted by the Dirac models. This means that the transport gap of zigzag antidots can be higher than that of similarly sized armchair antidots, which contradicts the behavior of the band gap \cite{brun2014electronic}. This is in good accordance with recent studies that showed that localized states in non-commensurate antidot lattices do not contribute to electronic transport \cite{lopata2010graphene}. Since the localized edge states typically lie beneath non-localized states, they will generally increase the transport gap. This also agrees with the results of Jippo \textit{et al.} \cite{jippo2011theoretical} who calculated the transport gap for irregularly shaped antidots. They showed that the transport gap generally increases with the length of consecutive zigzag regions in the antidot. The transport gap of armchair antidots is highly predictable, as it follows the simple result from the DMB very accurately, whereas the transport gap of zigzag antidots is much less predictable. Therefore, even though the transport gap can be higher in some cases for zigzag antidots compared to armchair antidots, it may be an advantage to use armchair antidots in an experimental setup.

\section{Conclusions}

We use a Green's tensor area integral equation method to solve the Dirac equation with a spatially varying mass term. In this way, we are able to calculate the scattering of Dirac electrons on arbitrary graphene antidot structures. 
We use this method to calculate the conductance of graphene antidot barriers with hexagonal antidots and compare them with results obtained with tight-binding. Our approach is much more general than previous attempts to use the Dirac equation to calculate scattering of Dirac electrons on antidots. The computational time of our Dirac model is scale invariant, which means that we are able to calculate properties of arbitrarily large structures. 
We show that our Dirac model is in excellent agreement with tight-binding for antidots with armchair edges. We also show that a simple Dirac mass barrier is able to predict the transport gap with high accuracy for antidots with armchair edges. Tight-binding predicts very localized edge states for large zigzag antidots, whereas the Dirac models do not. Therefore, the agreement between the Dirac models and tight-binding is generally poor when the barrier contains antidots with long zigzag edges. We show that the tight-binding transport gap for zigzag antidots is higher than for armchair antidots with equivalent size for some geometries, while it is lower for others. However, since the transport gap for armchair antidots is much more predictable, it may still be an advantage to use armchair antidots in an experimental setup. Furthermore, we show that a relatively narrow GAB, with only a few antidots in the unit cell, is sufficient to give rise to a transport gap.

\section*{Acknowledgments}
The authors gratefully acknowledge the financial support from the Center for Nanostructured Graphene (Project No. DNRF58) financed by the Danish National Research Foundation. We thank T. S{\o}ndergaard for helpful discussions on the Green's tensor area integral equation method and its numerical implementation.

\bibliography{literature}

\end{document}